# Rethinking Immersive Virtual Reality and Empathy

KEN JEN LEE, University of Waterloo, Canada
EDITH LAW, University of Waterloo, Canada

In this position paper, we aim to spark more discussions surrounding the use of empathy as the intended outcome of many studies on immersive virtual reality experiences. As a construct, empathy has many significant flaws that may lead to unintended and negative outcomes, going against our original goal of employing these technologies for the betterment of society. We highlight the possible advantages of designing for rational compassion instead, and propose alternative research directions and outcome measurements for immersive virtual reality that urgently warrant our attention.

**ACM Reference Format:**
Ken Jen Lee and Edith Law. 2021. Rethinking Immersive Virtual Reality and Empathy. In *arttech: Performance and Embodiment in Technology for Resilience and Mental Health Workshop in the 24th ACM Conference on Computer-Supported Cooperative Work and Social Computing (arttech@CSCW'21), October 23–24, 2021.* , 4 pages.

## 1 INTRODUCTION

Empathy is the "ability to understand and share the feelings of another" [7] and can be categorized into emotional (or affective) empathy and cognitive empathy. The former is one's emotional reactions to others', while the latter is the ability to recognize emotional states of others free of emotional arousal [14]. Many immersive virtual reality (IVR) experiences have, and successfully so, selected empathy as the main aim or intended outcome. For instance, Schutte and Stilinović found that an IVR of a documentary about a young girl living in a refugee camp elicited higher levels of empathy compared to watching the documentary in a two-dimensional format [20]. Similarly, Herrera et al. found that participants who experienced being homeless in a VR perspective-taking (VRPT) task are more likely to sign a measure supporting affordable housing four weeks after the task, compared to participants who merely received information through traditional and desktop-based forms of the task [10]. In another long-term study, Hasson et al. found that participants who underwent an IVR experience from an outgroup's point of view made them perceive the outgroup more favorably five months after the VR intervention [9]. Martingano et al. performed a meta-analysis of 43 IVR articles and found that IVR is capable of increasing emotional, but not cognitive, empathy [14], though van Loon et al.'s work found a VRPT task to be effective in increasing cognitive empathy for specific others [23].

Despite these demonstrated successes, we argue against the use of empathy as the intended outcome of IVR experiences in this position paper. As a construct, empathy has many significant flaws that could lead to unintended and negative outcomes, instead of improving wellbeing and encouraging human flourishing. We present a better alternative—rational compassion—and possible ways of quantifying it. We also motivate increasing the use of behavioral measures, and potential methods of measuring and avoiding the downsides of current empathy-based IVR research.

## 2 HOW COULD EMPATHY BE BAD?

Although empathy undoubtedly has many benefits, it is, in fact, not all good [1]. We will divide this discussion into the issues of using emotional empathy and cognitive empathy as the intended outcomes of IVR experiences.

Authors' addresses: Ken Jen Lee, kenjen.lee@uwaterloo.ca, University of Waterloo, Waterloo, Ontario, Canada; Edith Law, edith.law@uwaterloo.ca, University of Waterloo, Waterloo, Ontario, Canada.

**Emotional Empathy.** Although there is currently no consensus over IVR's general ability to increase cognitive empathy, IVR's ability to increase emotional empathy is much better understood and established [14]. However, mirroring others' emotions could lead to personal distress and subsequent hurtful behaviors. For instance, when empathizing with sexual assault victims, Martingano found that some participants blamed or distanced themselves from the victims because they mirrored victims' assumed shame [13]. In the long term, an overexposure to IVRs aimed at eliciting emotional empathy might also lead to empathic distress fatigue [11] or a decrease in capacity for empathy [18]. As such, even though emotional empathy could be associated with prosocial behaviors [10], it might be an unsustainable mechanism for encouraging such behaviors. Another downside of emotional empathy as an aim is that it is "not well-suited to support ethically correct decisions" due to the spotlight effect, where one's attention is narrowly focused on specific individuals [2]. This reflects current IVR studies that successfully made participants more empathetic only towards very specific social targets (e.g., [9, 10]). Moreover, the spotlight effect could be biased towards "those who are close to us, those who are similar to us, and those who we see as more attractive or vulnerable and less scary", and away from "those who are strange or different or frightening" [1]. As such, empathy could encourage quick side-taking judgements and polarization [3], and be manipulated towards certain individuals while leading to aversion towards others [2]. This means that at worst, emotional empathy could even lead us to actions attending to the suffering of a few that might result in terrible consequences for many more, since it is "particularly insensitive to consequences that apply statistically rather than to specific individuals" [1]. For a discussion of other types of negative acts that could result from emotional empathy, refer to Breithaupt's [2].

**Cognitive Empathy.** While emotional empathy could be "morally corrosive", cognitive empathy is more likely to just be "morally neutral" [1]. This neutrality is because cognitive empathy could be used for both good and bad motivations. On one side, cognitive empathy could "lead to prosocial behavioral intentions without emotional involvement" [13]. However, cognitive empathy could also be used for ill-intentioned manipulation of others' emotions (e.g., to incite violence), something Bubandt and Willerslev termed as "*tactical empathy*" [5]. Moreover, cognitive empathy, just like emotional empathy, is prone to the spotlight effect; IVR studies that managed to increase participants' cognitive empathy apply only to specific others [23].

This leads us to be aware of the dangers of empathy in the context of IVRs, particularly due to "*toxic empathy*" [16]. While Nakamura goes in-depth into the relation between recent trends towards exploring VR as the *ultimate empathy machine* and sociocultural history and factors in [16], we aim to highlight how IVR "mistakes point of view for embodied experience" [16], drawing parallels with white-to-Black racial passing, which only succeeded in providing an illusive sense of empathy and expertise, but "fails to bring about systemic or institutional racial change" [8, 21]. In other words, users might consume IVR experiences about specific marginalized groups or social issues just for the sake of wanting to feel empathy because 'empathic engagement might give [users] satisfaction" [1]. This satisfaction might even provide a sense of illusion that they now fully understood the issue or the marginalized group's experiences, and perhaps even consider themselves to be experts who have done their part in helping. In short, IVR experiences could become nothing more than a way for "feeling good about feeling bad" [16].

## 3 RATIONAL COMPASSION AS THE ALTERNATIVE

This position paper is certainly not the first work arguing against using empathy as the primary intended affective outcome of IVR studies aiming to increase overall morality and goodness towards others [15, 16]. As such, the more important part of this discussion should be about possible improvements and alternative outcome measures, a few of which are presented below.

First, we propose a shift of focus in IVR research from emotional empathy to rational compassion, a stand Bloom argues for in [1]. Rational compassion is "wanting to alleviate suffering and make the world a better place ... and a rational assessment of how best to do so" [1]. This idea of rational compassion is not new; Nassbaum put forward in an article more than two decades ago that "judgment that does not employ the intelligence of compassion in coming to grips with the significance of human suffering is blind and incomplete" [17]. Unlike emotional empathy, compassion itself does not require a mirroring of emotions [1]. Even when there is emotional resonance, compassion has an additional component: the ability to be nonjudgmental in accepting others' emotions, or a person's own emotional response to others' emotions [22]. As such, compassion could help build resilience instead of leading to personal distress, fatigue or burnout [19]. In other words, ideally, we argue that IVR research should not lead to significantly higher increases in personal distress, and instead facilitate rational compassion. However, how exactly can rational compassion be quantified? Compassion (and the closely related concept of self-compassion) have already been used as the targeted construct in IVR studies [4, 6]. But rational compassion also implies an ability to reason, defined as the act of explaining and justifying [12]. There are two possible ways of quantifying this. The first is to measure cognitive empathy itself. Given that compassion as a construct entails positive intentions and can include ambiguous and larger social targets [22], an increase in cognitive empathy alongside increases in compassion might be able to circumvent *tactical empathy*. The second possible way of measuring the *rational* aspect derives from the importance of prior knowledge on a person's reasoning process [24]. As a proxy, we could measure learning gains about the topic relevant to a specific IVR experience; we will revisit this later.

Second, we propose increasing the use of behavioral measures in IVR research. Part of the *toxic empathy* issue is that IVR might not translate into, and could even impede, real world structural changes [16]. Since the desire to act towards alleviating suffering is a core feature of compassion, unlike empathy [22], it would be interesting to assess whether designing for rational compassion could encourage prosocial and altruistic behaviors more effectively and sustainably. Herrera's et al.'s study is a great example of using behavioral measures (e.g., support for petition, willingness to donate) [10]. However, we could go one step further, and work with grassroots initiatives, NPOs or charities in participatory IVR design projects with quantifiable real world impacts serving as outcome measures, e.g., increases in donations, volunteers, or support for petitions and policies.

The popular view of VR as the *ultimate empathy machine* could also lead to *toxic empathy* since there is an implicit message that VR, as a technology and medium, is superior to other forms of learning and knowing about others' experiences, feelings and perspectives:

> "Here is the idea that you cannot trust marginalized people when they speak their own truth or describe their own suffering, but you have to experience it for yourself, through digital representation, to know that it is true." Nakamura in [16].

However, instead of replacing other ways of knowing, IVR can be an attractive gateway instead. Instead of measuring the effect of IVR as a standalone intervention, it could be measured instead in terms of its effectiveness as a way to increase people's subsequent learning gains about topics and issues that are otherwise neglected. For instance, after an IVR intervention, participants could be left in a room with various resources of more conventional formats (e.g., 2D videos, books, interviews, news articles etc.) about the similar topic, with the instruction that they could spend as long as they want looking through the resources. A positive result here would entail participants who experienced the IVR intervention spending significantly longer learning about the topic through conventional formats after the intervention, compared to control participants who did not experience the IVR intervention.

Finally, although compassion might not suffer from the spotlight effect because it can be felt for humanity at large [22], IVR research could also use more explicit measures of spotlight vision. For instance, if an IVR study is about a specific social target in a particular geographical area (e.g., homelessness in the U.S.), having behavioral measures corresponding to participants' interest and intentions of, or even actual, learning about other social targets (e.g., homelessness vs. ethnically marginalized groups) or similar social targets in other geographical areas (e.g., homelessness in a distant country) during or after an IVR intervention might provide more direct measurements of the spotlight effect.

Instead of hoping for a future where VR is the *ultimate empathy machine*, perhaps it is healthier and wiser to build a path towards a future where VR is just another medium for encouraging and facilitating rational compassion for human flourishing.